\documentclass[10pt]{iopart}
\usepackage[colorlinks=true, urlcolor=blue, linkcolor=blue, citecolor=blue, pdftex]{hyperref}
\usepackage{graphicx}% Include figure files
\usepackage{latexsym}
\usepackage{wasysym}
\usepackage{nicefrac}
\usepackage{iopams}
\usepackage{color}
\usepackage{dcolumn}% Align table columns on decimal point
\usepackage{bm}% bold math
\usepackage{braket}
\newcommand{\dagga}{{\phantom{\dagger}}}

\begin{document}

\title{Variational Monte Carlo study of stripes as a function of doping in  the $t-t'$ Hubbard model}
\author{Antonio Lechiara, Vito Marino, Luca F. Tocchio\footnote{Author to whom any correspondence should be addressed}}
\address{Institute for Condensed Matter Physics and Complex Systems, DISAT, Politecnico di Torino, I-10129 Torino, Italy}
\ead{luca.tocchio@polito.it}
\date{\today}

\begin{abstract}
We perform variational Monte Carlo simulations of the single-band Hubbard model on the square lattice with both nearest ($t$) and next-nearest ($t'$) neighbor hoppings. 
Our work investigates the consequences of increasing hole doping on the instauration of stripes and the behavior of the superconducting order parameter, 
with a discussion on how the two phenomena affect each other. We consider two different values of the next-nearest neighbor hopping parameter, 
that are appropriate for describing cuprate superconductors. We observe that stripes are the optimal state in a wide doping range; the stripe wavelength 
reduces at increasing doping, until stripes melt into a uniform state for large values of doping. Superconducting pair-pair correlations, indicating the presence of superconductivity, are always suppressed in the presence of stripes. Our results suggest that the phase diagram for
the single-band Hubbard model is dominated by stripes, with superconductivity being possible only in a narrow doping range between striped states and a nonsuperconducting metal.
\end{abstract}

\maketitle
\ioptwocol

\section{Introduction}

The concept of a stripe phase is one of the unconventional features  that emerged over the years when interpreting a 
broad range of experimental results on copper oxide superconductors. Once we allow holes to wander in an antiferromagnetic background, 
the creation of striped inhomogeneities can be a consequence. The reason for the self-organization of local inhomogeneities can be found 
in the competition between the tendency of the electrons to cluster in antiferromagnetic regions, hence producing a short-range tendency to phase separation, 
and the long-range Coulomb interaction that instead frustrates it~\cite{emery1993,castellani1995,hellberg1997,emery1999}. 
Striped states indeed constitute the best compromise between these 
competing phenomena and allow the doping holes to be delocalized along linear stripes. 

From an experimental point of view, the first evidence of stripes comes 
from a neutron scattering study on a single crystal of  La$_{1.48}$Nd$_{0.4}$Sr$_{0.12}$CuO$_4$~\cite{tranquada1995}. 
Since then, a variety of experimental probes, based on neutron scattering, x-ray scattering 
and scanning tunneling microscopy, pointed to the presence of spin and charge orders~\cite{bianconi1996,hoffman2002,howald2003,ghiringhelli2012,tranquada2013,comin2016}. Nuclear magnetic resonance (NMR) is also a valuable tool to study charge and spin modulations in cuprates. In particular, NMR measures indicated that cuprates that are not La-based may exhibit charge order without spin order~\cite{wu2011,zhou2024}. Moreover, NMR has been employed to investigate the nature of the pseudogap critical point when superconductivity is suppressed~\cite{frachet2020}.

The simplest model that has been considered to reproduce the essential features of the cuprates' phase diagram is the single-band Hubbard model, 
where only the $d_{x^2-y^2}$ orbital of Cu atoms is retained and the impact of oxygen atoms is neglected. However, despite its simplicity,
obtaining accurate approximations for the ground state and for low-energy excitations is far from being trivial and several 
states, very close in energy, have been proposed, obtaining different conclusions from different numerical and analytical methods~\cite{qin2022}. 
The model is reported here:
\begin{eqnarray}\label{eq:Hubbard}
\fl {\cal H} = -t \sum_{\langle R,R^\prime \rangle,\sigma} c^\dagger_{R,\sigma} c^\dagga_{R^\prime,\sigma} \nonumber \\
-t^\prime \sum_{\langle\langle R,R^\prime \rangle\rangle,\sigma} c^\dagger_{R,\sigma} c^\dagga_{R^\prime,\sigma} +\textrm{H.c.} + U \sum_{R} n_{R,\uparrow}n_{R,\downarrow},
\end{eqnarray}
where $c^\dagger_{R,\sigma}$ ($c^\dagga_{R,\sigma}$) creates (destroys) an electron with spin $\sigma$ on site $R$ and 
$n_{R,\sigma}=c^\dagger_{R,\sigma} c^\dagga_{R,\sigma}$ is the electron density per spin $\sigma$ on site $R$. In the following, we indicate the 
coordinates of the sites with ${\bf R}=(x,y)$. It is important that the Hubbard model includes not only the nearest neighbor hopping $t$ and 
the on-site electron-electron repulsion $U$, but also the next-nearest neighbor hopping $t'$ that has been shown to be a relevant feature in all cuprates, 
as it constitutes an essential material-dependent parameter, with $t'/t<0$~\cite{pavarini2001}. 
The electron density is given by $n=N/L$, where $N$ is the number of electrons and $L$ is the total 
number of sites. The hole doping is defined as $\delta=1-n$.

In the $t'=0$ case, the presence of stripes in the Hubbard model originates from density-matrix renormalization group (DMRG) 
studies on 6-leg ladders~\cite{white2003,hager2005} and from further works supporting the idea that charge and spin inhomogeneities may 
pervade the phase diagram of the Hubbard model~\cite{chang2010,zheng2016}. Charge modulations have been also proposed to be present and to possibly enhance superconductivity by the Dynamic Cluster Approximation and by determinant Quantum Nonte Carlo~\cite{maier2010,mondaini2012}. Later, a work which combined a
variety of numerical techniques~\cite{zheng2017}, focused on the representative doping $\delta=1/8$ at $U/t=8$, settling that the lowest-energy stripe 
is a bond-centered one with periodicity $\lambda=8$ (named the stripe wavelength) in the charge sector and $2\lambda=16$ 
in the spin sector. As a consequence, the enlarged unit cell of length $\lambda$ contains, on average, one hole, as obtained by previous Hartree-Fock 
calculations~\cite{poilblanc1989,zaanen1989,machida1989,schulz1989}. Electron pairing was not found in this case and also further studies 
highlighted the absence of superconductivity at doping $\delta=1/8$, the system being possibly an insulator~\cite{tocchio2019b,sorella2023,qin2020}.
Away from doping $\delta=1/8$, stripes have been proposed to be stable by different numerical methods, possibly coexisting with 
superconductivity~\cite{tocchio2019b,vanhala2018,darmawan2018,xu2022}. Recently, an accurate variational auxiliary field quantum Monte Carlo study 
proposed a phase diagram where stripe phases with no superconductivity are present close to half filling while a superconductive region emerges 
around $\delta\sim 0.2$~\cite{sorella2023}. 

In the $t'\ne 0$ case, stripes emerge already in the Hartree-Fock approximation, for $t'/t <0$~\cite{scholle2023}. Then, their presence is confirmed by different numerical methods, which agree in observing a reduction of the stripe wavelength when $t'/t<0$, while 
the stripe wavelength increases when $t'/t>0$~\cite{huang2018,ido2018,ponsioen2019,marino2022}. 
There is consensus on the absence of superconductivity at $\delta=1/8$, while different outcomes
on the existence of superconductivity are reported when other dopings are considered. 
The debate is still open since, recently, a combined study based on DMRG and AFQMC indicates that partially 
filled stripes coexist with superconductivity in a large doping range of the $t-t'$ Hubbard model~\cite{xu2024}, 
while a DMRG study on 6-leg cylinders suggests that superconducting correlations decay exponentially for $t'/t<0$~\cite{jiang2024}.

Alternatively, it has been suggested that superconductivity can be enhanced when models less simplified than the Hubbard one are taken into account.
For instance, superconductivity can be recovered, without clear long-range stripe order, with an \emph{ab-initio} approach 
that highlights the presence of the realistic off-site interactions~\cite{schmid2023}. Moreover, the three-band Hubbard (or Emery) model 
has been proposed as a way to enhance superconductivity~\cite{jiang2023a,ponsioen2023}, while hopping modulation in a stripe-like manner 
has been suggested to enhance superconductivity even in the pure Hubbard model~\cite{jiang2022}. Fluctuating stripes have been also proposed to coexist with superconductivity, at difference with the static ones, in the attractive Hubbard model~\cite{sun2024}.

In this paper, we employ the variational Monte Carlo method with backflow correlations to investigate the effect of doping 
on stripes and superconductivity in the $t-t'$ Hubbard model, for $t'/t=-0.25$ and $t'/t=-0.40$ at $U/t=8$. Our simulations are performed 
on 6-leg ladders, with $L_x$ rungs, the total number of sites being $L=L_x \times 6$. This geometry has been employed in DMRG calculations 
and is expected to capture the properties of truly two-dimensional clusters~\cite{zheng2017}, 
while it allows us to accommodate long stripes along the rungs. First of all, we show that bond-centered and site-centered stripes 
have similar energies, the only relevant quantity being the stripe wavelength $\lambda$. Then, we observe that stripes are the optimal state in a wide doping range: 
The stripe wavelength reduces at increasing doping, until stripes melt into a uniform state for large values of the doping. We show that 
superconducting correlations are always suppressed in the presence of stripes, regardless of their insulating or metallic character. 
Instead, when the stripes melt into a uniform state, a narrow region of superconductivity is observed, around $\delta\sim 0.30$ for $t'/t=-0.25$ 
and around $\delta\sim 0.21$ for $t'/t=-0.40$. We also report on the effect of the next-nearest neighbor hopping, showing that a larger value of $|t'/t|$ 
induces a faster disruption of both stripes and superconductivity as a function of doping. The results discussed here are based on the Master's thesis of the first author of the paper.~\cite{lechiara2023}.

\section{Method}\label{method}

Our numerical results are obtained with the variational Monte Carlo method (VMC), which is based on the definition of correlated variational wave functions, whose 
physical properties can be evaluated within a Monte Carlo scheme~\cite{beccabook}. In particular, electron-electron correlation is inserted by means of a density-density 
Jastrow factor~\cite{capello2005,capello2006} on top of a Slater determinant or a Bardeen-Cooper-Schrieffer (BCS) state. In addition, backflow correlations,
as described in \cite{tocchio2008,tocchio2011}, are implemented; the latter ingredient is important to get accurate variational states. 

The wave function is defined by:
\begin{equation}\label{eq:wf_SL}
|\Psi\rangle={\cal J}_d |\Phi_0\rangle\,,
\end{equation}
where ${\cal J}_d$ is the density-density Jastrow factor and $|\Phi_0\rangle$ is a state that is constructed from the ground state of an auxiliary 
noninteracting Hamiltonian by applying backflow correlations~\cite{tocchio2008,tocchio2011}. The variational wave functions described below are similar to the ones we built to study the parameter regimes discussed in Refs.~\cite{tocchio2019b,marino2022}.

The Jastrow factor is given by
\begin{equation}\label{eq:jastrowd}
{\cal J}_d = \exp \left ( -\frac{1}{2} \sum_{R,R'} v_{R,R^\prime} n_{R} n_{R^\prime} \right )\,,
\end{equation}
where $n_{R}= \sum_{\sigma} n_{R,\sigma}$ is the electron density on site $R$ and $v_{R,R^\prime}$ are pseudopotentials that are optimized for 
every independent distance $|{\bf R}-{\bf R}^\prime|$ of the lattice. 
 
\begin{figure}
\includegraphics[width=0.5\textwidth]{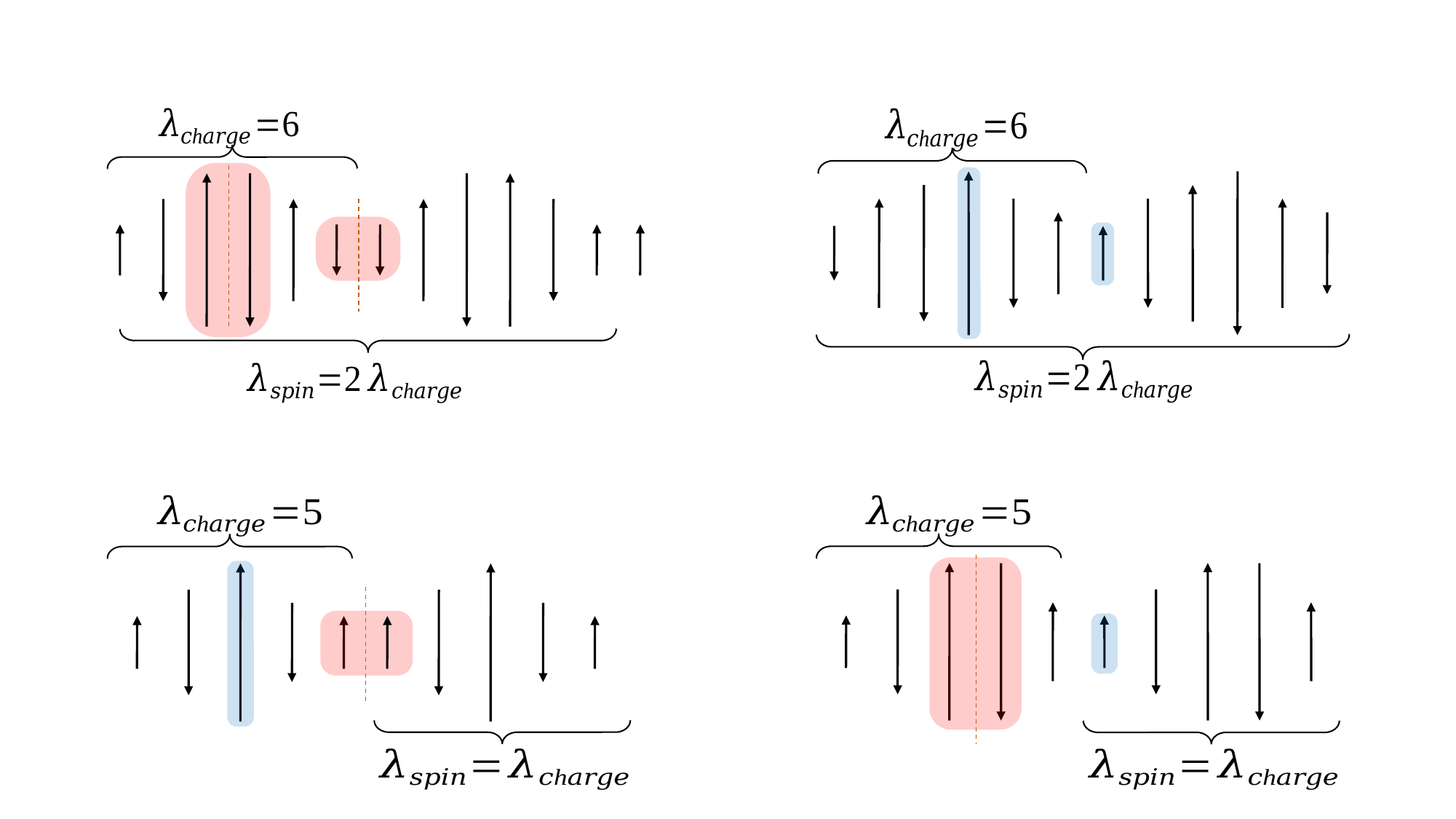}
\centering
\caption{Top left panel: spin modulation of a bond-centered stripe with even wavelength. The minimum and the maximum of the local magnetic momenta are located on bonds, as highlighted by the red rectangles. Top right panel: spin modulation of a site-centered stripe with even wavelength. The minimum and the maximum of the local magnetic momenta are located on sites, as highlighted by the blue rectangles. Bottom panels: spin modulations of a stripe with odd wavelength; in the left panel the modulation is bond-centered where the local magnetic moment is minimal (red rectangle) and site-centered where the local magnetic moment is maximal (blue rectangle); in the right panel the situation is reversed.}
\label{fig:stripes_sketch}
\end{figure}

In the case of stripe states, the auxiliary noninteracting Hamiltonian is defined as:
\begin{equation}\label{eq:auxham_stripes}
 {\cal H}_{\rm{aux}} = {\cal H}_{0} + {\cal H}_{\rm{charge}} + {\cal H}_{\rm{spin}} + {\cal H}_{\rm{BCS}}.
\end{equation}
The first one defines the kinetic energy of the electrons:
\begin{equation}\label{eq:ham0}
{\cal H}_{0} = -t \sum_{\langle R,R^\prime \rangle,\sigma} c^\dagger_{R,\sigma} c^\dagga_{R^\prime,\sigma} 
-\tilde{t}^\prime \sum_{\langle\langle R,R^\prime \rangle\rangle,\sigma} c^\dagger_{R,\sigma} c^\dagga_{R^\prime,\sigma} 
+\textrm{H.c.},
\end{equation}
where the value of the nearest neighbor hopping parameter $t$ is fixed to be equal to the one in the Hubbard Hamiltonian of Eq.~(\ref{eq:Hubbard}), 
in order to set the energy scale. Then, the second and third terms describe linear stripes along the $x$ direction that can be either bond-centered or site-centered:
\begin{equation}\label{eq:charge_mod}
{\cal H}_{\rm{charge}} = \Delta_{\rm{c}} \sum_{R} \cos \left [ Q \left ( x-x_0 \right ) \right ] 
\left ( c^{\dagger}_{R,\uparrow} c^{\phantom{\dagger}}_{R,\uparrow} + 
c^{\dagger}_{R,\downarrow} c^{\phantom{\dagger}}_{R,\downarrow} \right ),
\end{equation}
and
\begin{eqnarray}\label{eq:spin_mod}
\fl {\cal H}_{\rm{spin}} = \Delta_{\rm{s}} \sum_{R}(-1)^{x+y} \sin \left [ \frac{Q}{2} \left ( x-x_0 \right ) \right ] \times \nonumber \\
\times \left ( c^{\dagger}_{R,\uparrow} c^{\phantom{\dagger}}_{R,\uparrow} 
- c^{\dagger}_{R,\downarrow} c^{\phantom{\dagger}}_{R,\downarrow} \right ).
\end{eqnarray}
If $x_0=1/2$ the stripes are symmetric with respect to the bond halfway in between two neighboring lattice sites, hence they are called \textit{bond centered}. 
Conversely, for $x_0=0$, the stripes are called \textit{site centered}, as the symmetry axis lies exactly on a lattice site. 
The periodicity of the charge modulation in both cases is given by $\lambda=2\pi/Q$. On the other hand, the spin modulation 
has a $\pi$-phase shift across the sites with maximal hole density, resulting in a spin modulation 
of $2\lambda=4\pi/Q$ when $\lambda$ is even and $2\pi/Q$ when $\lambda$ is odd. 
The spin modulation along the $y$ direction is assumed to have N\'eel order in all cases. For clarity, we report in Fig.~\ref{fig:stripes_sketch} a sketch of the spin modulation along the $x$ direction, for bond- and site-centered stripes, in the case of even and odd wavelengths $\lambda$; the length of the arrows is proportional to the size of the local magnetic moments. 
The effect of the $\pi$-shift is clearly visible in the figures. 
The last term in Eq.~(\ref{eq:auxham_stripes}) introduces BCS electron pairing:
\begin{eqnarray}\label{eq:bcs}
\fl {\cal H}_{\rm{BCS}} = \sum_{R,\eta=x,y} \Delta_{R,R+\eta} (c^{\dagger}_{R,\uparrow} c^{\dagger}_{R+\eta,\downarrow} - 
c^{\dagger}_{R,\downarrow} c^{\dagger}_{R+\eta,\uparrow}) + \textrm{H.c.} \nonumber \\
- \mu \sum_{R,\sigma} c^{\dagger}_{R,\sigma} c^{\phantom{\dagger}}_{R,\sigma},
\end{eqnarray}
where the pairing amplitude is modulated in real space:
\begin{eqnarray}\label{eq:modulated_pairing}
 &\Delta_{R,R+x} = \Delta_x \left | \cos \left [ \frac{Q}{2}(x+\frac{1}{2}-x_0) \right ] \right | \nonumber \\
 &\Delta_{R,R+y} = -\Delta_y \left | \cos \left [ \frac{Q}{2} \left ( x-x_0 \right ) \right ] \right |.
\end{eqnarray}
This modulation has been named ``in phase'' in \cite{himeda2002}. A chemical potential $\mu$ is also included in ${\cal H}_{\rm{BCS}}$.

%%%%%%%%%%%%%%%%%%%%%%%%%%%%%%%%%%%%%%%%%%%%%%%%
%%%%%%%%%%%%%%%%%%%%%%%%%%%%%%%%%%%%%%%%%%%%%%%%%

In the case of uniform states, the auxiliary Hamiltonian is defined as:
\begin{equation}\label{eq:auxham_uniform}
 {\cal H}_{\rm{aux}} = {\cal H}_{0} + {\cal H}_{\rm{BCS}} +  {\cal H}_{\rm{AF}}.
\end{equation}
The kinetic term ${\cal H}_0$ is defined as for striped states. The BCS electron pairing is now defined without modulation 
in real space, i.e. $\Delta_{R,R+x} = \Delta_x$ and $\Delta_{R,R+y} = -\Delta_y$, for each site $R$.
In addition, a standard N\'eel order with pitch vector ${\bf K}=(\pi,\pi)$ can be considered in the uniform state:
\begin{equation}\label{eq:Neel}
{\cal H}_{\rm{AF}} = \Delta_{\rm{AF}} \sum_{R} (-1)^{x+y} \left ( c^{\dagger}_{R,\uparrow} c^{\phantom{\dagger}}_{R,\uparrow} -
c^{\dagger}_{R,\downarrow} c^{\phantom{\dagger}}_{R,\downarrow} \right ).
\end{equation}

The auxiliary Hamiltonians of Eqs.~(\ref{eq:auxham_stripes}) and (\ref{eq:auxham_uniform}) can be diagonalized by standard methods. 
Its ground state is then constructed. On top of it, backflow correlations are inserted to define $|\Phi_0\rangle$ of Eq.~(\ref{eq:wf_SL}), 
following our previous works~\cite{tocchio2008,tocchio2011}.

All the parameters in the trial wave function are optimized with the stochastic reconfiguration method~\cite{sorella2005}, in order to minimize the variational energy.
In particular, for striped states, we fix a given stripe wavelength
$\lambda$ and optimize $\Delta_x$, $\Delta_y$, $\mu$, $\Delta_{c}$, $\Delta_{s}$, and $\tilde{t}^\prime$ (as well as all the 
pseudopotentials in the Jastrow factor and the backflow parameters). For a uniform state, we do not have $\Delta_{c}$ and $\Delta_{s}$ 
as parameters and we optimize instead $\Delta_{\rm{AF}}$.
Once the energy and all the parameters converge to stable values, the optimization run can be
concluded. The values of the parameters are fixed to their averages and a run at fixed parameters is performed to compute the quantum averages needed for the
correlation function or the superconducting order parameter.

As already discussed in \cite{tocchio2019b,marino2022}, finite values of $\Delta_c$ and $\Delta_s$ 
lead effectively to charge and spin modulations, as signaled by a peak (diverging in the thermodynamic limit) at a given ${\bf Q}$ vector
in the static spin and charge structure factors. 
%\begin{equation}\label{eq:sqsq}
%S({\bf q}) = \frac{1}{L} \sum_{R,R^\prime}\langle S^z_{R} S^z_{R^\prime} \rangle e^{i{\bf q}\cdot({\bf R}-{\bf R^\prime})},
%\end{equation}
%where $\langle \dots \rangle$ indicates the expectation value over the variational wave function and
%$S^z_{R}=1/2(c^{\dagger}_{R,\uparrow} c^{\phantom{\dagger}}_{R,\uparrow} - c^{\dagger}_{R,\downarrow} c^{\phantom{\dagger}}_{R,\downarrow})$. 
In order to assess the metallic or the insulating nature of the ground state, we can investigate the small-$q$ behavior of the static charge structure factor
$N({\bf q})$, defined as:
\begin{equation}\label{eq:nqnq}
N({\bf q}) = \frac{1}{L} \sum_{R,R^\prime}\langle n_{R} n_{R^\prime} \rangle e^{i{\bf q}\cdot({\bf R}-{\bf R^\prime})},
\end{equation}
where $\langle \dots \rangle$ indicates the expectation value over the variational wave function. Indeed, charge 
excitations are gapless when $N({\bf q}) \propto |{\bf q}|$ for $|{\bf q}| \to 0$, while a charge gap is present whenever 
$N({\bf q}) \propto |{\bf q}|^2$ for $|{\bf q}| \to 0$~\cite{tocchio2011,feynman1954}. 

The possible existence of superconductivity is investigated by computing correlation functions between Cooper pairs at distance $r$ along the $x$ direction. In particular,
we can consider pairs along the $y$ direction, so that:
\begin{equation}\label{eq:pairing}
D(r) = \frac{1}{L} \sum_{R} \langle P^\dagga_{R} P^\dag_{R+rx} \rangle,
\end{equation}
where $P^\dagga_{R} = c^\dagga_{R+y,\downarrow} c^\dagga_{R,\uparrow} - c^\dagga_{R+y,\uparrow} c^\dagga_{R,\downarrow}$ destroys
two electrons at nearest-neighbor sites (along $y$). Then, superconductivity exists whenever $D(r)$ does not decay to zero at large values of $r$. 

Our simulations are performed on ladders with
$L=L_x \times 6$ sites and periodic boundary conditions along both the $x$ and the $y$ directions. In order to fit charge and spin patterns in 
the cluster, we take $L_x=2k\lambda$ (with $k$ integer). Some analysis of the dependence of numerical results on $L_y$ and $L_x$ can be found in Refs.~\cite{tocchio2019b,marino2022}.

\section{Results}

In this section, we study the instauration of superconductivity and stripes of different wavelength $\lambda$ when changing the hole doping $\delta$.
We consider two typical values of the hopping parameter for cuprates ($t'/t = -0.25$ and $t'/t = -0.4$) in order to see how the value of $t'/t$ affects the stripe order. The on-site Coulomb
repulsion $U/t = 8$, kept fixed throughout the simulations, is chosen to ensure strong enough correlations. Indeed, in \cite{marino2022}, 
it is shown that for smaller values of $U/t$, such as $U/t \lesssim 4$, the striped wave functions are not stable and converge to
the uniform state with vanishing parameters $\Delta_c$ and $\Delta_s$. 

We consider both commensurate and incommensurate doping values. By "commensurate" doping we refer to the introduction of an integer
number of holes every $1/\delta$ lattice sites; conversely, this number is noninteger for "incommensurate" doping values.

The optimal variational state is found by comparing, for each considered value of $t'/t$ and $\delta$, the variational energies 
of the striped states for various $\lambda$ and of the uniform state. The optimal state is then the one with the lowest energy. 

\subsection{$t'/t=-0.25$}

We start by considering the case $t'/t = -0.25$. The energy per site, in units
of $t$, as a function of $\delta$ is reported in Table~\ref{tab:energies1}. Here, we compare the energy
for the best striped state $E_{\rm{stripe}}/t$ with that of the uniform state $E_{\rm{uniform}}/t$ for a
broad range of doping values. The striped state is almost always energetically favorable. As $\delta$ increases,
the wavelength $\lambda$ decreases more and more until, at doping $\delta = 1/3$, 
the striped state and the uniform state become energetically indistinguishable.
When discussing the behavior of the gap parameters, we will show
that this effectively corresponds to a melting of the stripe.

%%%%%%%%%%%%%%%%%%%%%%%%%%%%%%%%%%%%%%%%%%%%%%%%%%%%%%%%%%%%%%%%%%%%%
\begin{table}
\caption{\label{tab:energies1}
Energy per site (in units of $t$) for the best striped state $E_{\rm{stripe}}$ and the uniform state $E_{\rm{uniform}}$, 
along with their relative difference $\Delta E=E_{\rm{stripe}}-E_{\rm{uniform}}$, as a function of $\delta$ for $t^\prime/t=-0.25$. 
Data are shown for $L_x=48$ at dopings $1/12, 1/6, 1/4, 1/3$, for $L_x=45$ at doping 1/5, for $L_x=40$ at doping 1/8, and for $L_x=70$ at doping 1/10. These values of $L_x$ are chosen to accommodate the selected dopings, fit charge and spin patterns of the stripes, and be large enough to have limited size effects. The error bar on the energy, of the order of  $10^{-4}t$, takes into account the weak lattice size dependence of the energies.}
\begin{center}
\begin{tabular}{@{}cccc}
\br
$\delta$   &  $E_{\rm{stripe}}$ & $E_{\rm{uniform}}$ & $\Delta E$ \\
\mr
 1/12 & -0.6646 ($\lambda=8$) & -0.6572 & -0.0074 \\ 
 1/10 & -0.6920 ($\lambda=7$) & -0.6842 & -0.0078 \\
 1/8 & -0.7322 ($\lambda=5$) & -0.7238 & -0.0084 \\
 1/6 & -0.7936 ($\lambda=4$) & -0.7847 & -0.0088 \\
 1/5 & -0.8281 ($\lambda=3$) & -0.8258 & -0.0023 \\
 1/4 & -0.8749 ($\lambda=3$) & -0.8727 & -0.0021 \\
 1/3 & -0.9197 ($\lambda=3$) & -0.9197 & 0 \\
\br
\end{tabular}
\end{center}
\end{table}
%%%%%%%%%%%%%%%%%%%%%%%%%%%%%%%%%%%%%%%%%%%%%%%%%%%%%%%%%%%%%%%%%%%

\begin{figure}
\includegraphics[width=0.5\textwidth]{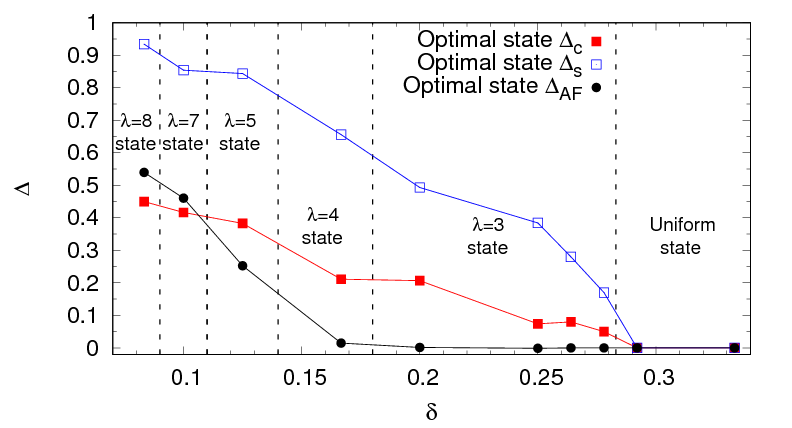}
\centering
\caption{Behavior of $\Delta_{c}$ (full squares) and $\Delta_{s}$ (empty squares) 
for the best striped state and $\Delta_{\rm{AF}}$ (circles) for the uniform state, 
as a function of $\delta$ for $t^\prime/t=-0.25$. The best striped state for each doping is shown in figure. The error bars are smaller than the symbol size.}
\label{fig: gap behav}
\end{figure}

In Sec.~\ref{method}, we have introduced the gap parameters $\Delta_c$, $\Delta_s$ and $\Delta_{\rm{AF}}$ related to the "strength" of the charge, 
spin, and N\'eel order, respectively. In this section we proceed by looking at their behavior as the hole-doping increases in the case $t^\prime/t=-0.25$.
Their values are plotted in Fig.~\ref{fig: gap behav}. For small doping, the striped state is well established and indeed $\Delta_{c}$ and $\Delta_{s}$ are finite. 
This corresponds to a well-defined order in both charge and spin. Also the uniform state, despite not being the optimal one, 
is able to develop N\'eel antiferromagnetism in the underdoped regime, as indicated by a finite $\Delta_{\rm{AF}}$. 

As $\delta$ increases, we see that all these parameters decrease monotonically until, 
at large values of $\delta$ they become much smaller and eventually negligible. This corresponds, for the striped states, to the absence of any order: 
the stripe "melts" and effectively reduces to the uniform state. Hence the degeneracy in energy pointed out for $\delta=1/3$. 
Our results confirm the shrinking of the stripes at increasing doping, i.e., shorter modulations are favored when more holes are present in the system.

%%%%%%%%%%%%%%%%%%%%%%%%%%%%%%%%%%%%%%%%%%%%%%%%
%%%%%%%%%%%%%%%%%%%%%%%%%%%%%%%%%%%%%%%%%%%%%%%%%

\begin{figure}
\centering
\includegraphics[width=0.5\textwidth]{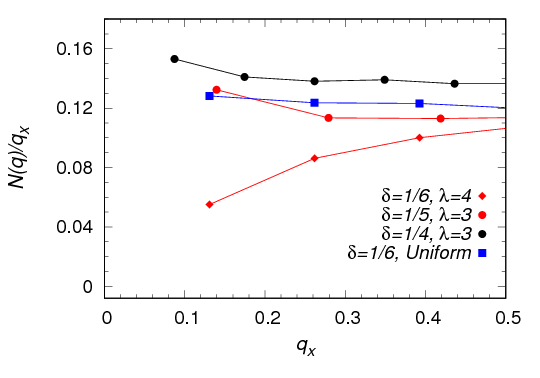}
\caption{Static structure factor (divided by $q_x$) $N(\textbf{q})/q_x$ as a function of $q_x$ with $q_y=0$. 
Data are reported at $t'/t=-0.25$, for the (nonoptimal) uniform state at doping $\delta=1/6$ for $L_x=48$ (squares), and for the optimal striped states at dopings $\delta=1/6$ (diamonds), 1/5 (red circles), 1/4 (black circles) (for $L_x=48,45,72$, respectively). The error bars are smaller than the symbol size.}
\label{fig: met o ins}
\end{figure}

The metallic or insulating behavior of the optimal state can be assessed from the small-$q$ behavior of the static structure factor $N(\textbf{q})$, 
see Eq.~(\ref{eq:nqnq}). In particular, we plot the quantity $N(q_x,0)/q_x$ at small $q_x$, as shown in Fig.~\ref{fig: met o ins}.

%%%%%%%%%%%%%%%%%%%%%%%%%%%%%%%%%%%%%%%%%%%%%%%%%%%%%%%%%%%
%%%%%%%%%%%%%%%%%%%%%%%%%%%%%%%%%%%%%%%%%%%%%%%%%%%%%%

As a reference, we used a uniform state (squares), which is known to be metallic 
(except at half-filling, when each site is occupied by one electron and the Coulomb repulsion prevents them from moving freely) 
even though it has a higher variational energy. 
We observe that, for the striped state at $\delta=1/6$ (diamonds), $N(q_x,0)/q_x$ clearly tends to zero, 
compatibly with an insulating behavior. On the other hand, for all the other striped states at $\delta=1/5$ and $\delta=1/4$ (circles), 
$N(q_x,0)/q_x$ tends to a finite value indicating that these states are metallic. 

\begin{figure}
\includegraphics[width=0.5\textwidth]{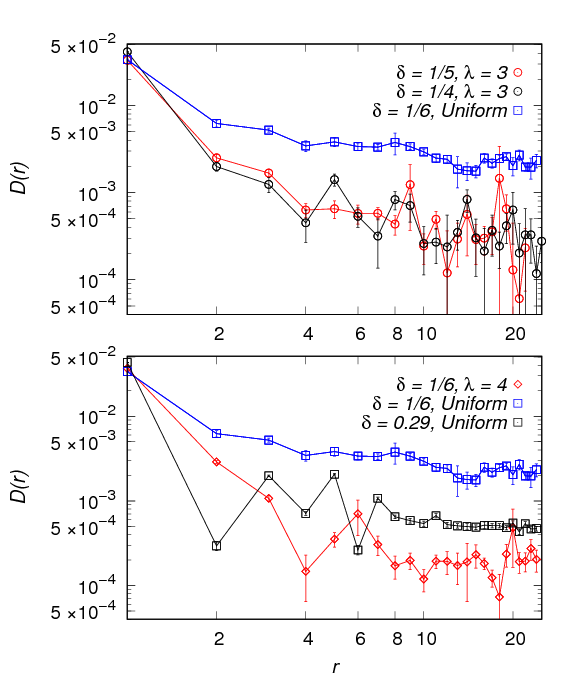}
\centering
\caption{Pair-pair correlations $D(r)$ as a function of $r$ on a log-log scale, at $t'/t=-0.25$. Upper panel: Data are reported for the optimal stripe states at different hole-dopings $\delta$ (circles) and for the (nonoptimal) uniform state at doping $\delta=1/6$ (blue squares).
Lower panel: Data are reported for the uniform state at the critical doping $\delta_c=0.29$ (black squares) along with the (nonoptimal) uniform 
superconducting state at $\delta=1/6$ (blue squares) and the nonsuperconducting striped state with $\lambda=4$ at $\delta=1/6$ (diamonds).}
\label{fig:corrSC}
\end{figure}

Finally, we address the coexistence of superconductivity and stripe order, by computing the superconducting order parameter of Eq.~(\ref{eq:pairing}). In Fig.~\ref{fig:corrSC}, we compare the uniform (but not optimal) state at $\delta=1/6$ (blue squares), taken as a reference, with some optimal striped states. All the striped states show strongly suppressed pair-pair correlations with respect to the uniform case. The stripes at $\delta=1/5$ and $\delta=1/4 $(circles), despite having a metallic character, exhibit a suppression in $D(r)$ similar to that of the insulating stripe at $\delta=1/6$ (diamonds). This supports the idea that the stripe order disrupts superconductivity, no matter their metallic or insulating character. 

%%%%%%%%%%%%%%%%%%%%%%%%%%%%%%%%%%%%%%%%%%%%5

Since stripes are found to compete with superconductivity, we investigate then whether there is a region where 
hole-doping is strong enough to restore the uniform state but not too strong to suppress superconductivity. 
In order to answer this question, we look for the first value of $\delta$ at which the uniform state becomes energetically favorable and compute the pair-pair correlations.  
For $t^\prime/t=-0.25$, as discussed in Table \ref{tab:energies1}, the optimal state at $\delta=1/4$ is a stripe of wavelength $\lambda=3$ 
while at $\delta=1/3$ we have already reached the uniform state. We then study incommensurate values of $\delta$ in the range $\left[\frac{1}{4},\frac{1}{3}\right]$. 
Since the wavelengths of the stripes decrease at increasing doping, it is sufficient to compare the striped state with $\lambda=3$ 
and the uniform state in this doping regime. Their variational energies are presented in Table \ref{table:incomm025}.

%%%%%%%%%%%%%%%%%%%%%%%%%%%%%%%%%%%%%%%%%%%%%%%%%%%%%%%%%%%%%%%%
\begin{table}
\caption{\label{table:incomm025} Energy per site (in units of $t$) for the best striped state $E_{\rm{stripe}}$ with $\lambda=3$ and the uniform state $E_{\rm{uniform}}$, 
along with their relative difference $\Delta E=E_{\rm{stripe}}-E_{\rm{uniform}}$, as a function of incommensurate $\delta$ for $t^\prime/t=-0.25$. 
Data are shown for $L_x=48$ for all the stripes and the uniform state. The error bar on the energy is always smaller than $10^{-4}t$.}
\begin{center}
\begin{tabular}{@{}cccc} 
 \br
 $\delta$ & $E_{\rm{stripe}}$ & $E_{\rm{uniform}}$ & $\Delta_E$ \\  
 \mr
 0.26 & -0.8799  & -0.8784 & -0.0015 \\
 0.27 & -0.8891  & -0.8885 & -0.0006 \\
 0.28 & -0.8933  & -0.8932 & -0.0001 \\
 0.29 & -0.9015  & -0.9016 & 0.0001 \\
 0.31 & -0.9086  & -0.9087 & 0.0001 \\
 \br
\end{tabular}
\end{center}
\end{table}
%%%%%%%%%%%%%%%%%%%%%%%%%%%%%%%%%%%%%%%%%%%%%%%%%%%%%%%%%%%

 We can identify as the critical doping, 
 the value $\delta_c=0.29$, where the optimal parameters of the stripe state 
$\Delta_c$ and $\Delta_s$ vanish. In Fig.~\ref{fig:corrSC}, lower panel, the 
pair-pair correlations for this state (black squares) are plotted next to the uniform, but not 
optimal, superconducting state at $\delta=1/6$ (blue squares) and the nonsuperconducting striped state at $\delta=1/6$ with $\lambda=4$ (diamonds), for comparison. 
 In this "intermediate" state, superconductivity is suppressed with respect to the uniform state at smaller doping, due to an already strong hole-doping, 
 but is still present, at difference with the striped state.

\subsection{$t'/t=-0.4$}

The second set of simulations involved the same search for the optimal state but at a larger value of $|t^\prime/t|$, namely $t^\prime/t=-0.4$. 
The main effect of a larger value of $|t^\prime/t|$ is to suppress the stripe pattern and makes the optimal state converge to the uniform one faster. 
Indeed, while at $\delta=1/5$ the optimal state is a stripe with $\lambda=3$, already at $\delta=1/4$ the striped state is no longer favorable 
and the uniform state is the optimal one. We can ascribe this behavior to the larger degree of frustration that is present for a larger value of $|t'/t|$.

\begin{figure}
\includegraphics[width=0.5\textwidth]{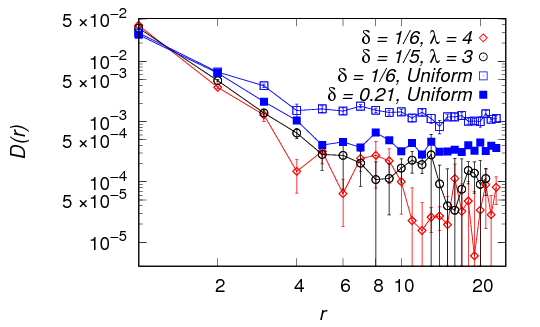}
\centering
\caption{Pair-pair correlations $D(r)$ as a function of $r$ on a log-log scale. Data are reported at $t'/t=-0.4$ for the (nonoptimal) uniform state at doping $\delta=1/6$ (empty squares) and 
for the optimal striped states at dopings $\delta=1/6$ (diamonds), 1/5 (circles), as well as for the uniform state at the critical doping $\delta_c=0.21$ (full squares).}
\label{fig: corrSC04_trans}
\end{figure}

We investigate then whether the suppression of the stripes at a lower concentration of holes might be associated to the presence of 
stronger superconducting correlations around the critical doping $\delta_c$. Following the same reasoning as before, the doping at which the uniform state prevails 
is for $\delta$ in the range $\left[\frac{1}{5},\frac{1}{4}\right]$. In particular, stripes are suppressed already at $\delta_c=0.21$. In Fig.~\ref{fig: corrSC04_trans}, we show the pair-pair correlations for the case $t^\prime/t=-0.4$ 
for different values of doping. Again, we can see how correlations for the uniform state at the critical doping (full squares)
are suppressed with respect to the uniform but not optimal, superconducting state at $\delta=1/6$ (empty squares), 
but still stronger than in the cases where stripe order is established.

%%%%%%%%%%%%%%%%%%%%%%%%%%%%%%%%%%%%%%%%%%%%%%%%%%%

\begin{figure}
\includegraphics[width=0.5\textwidth]{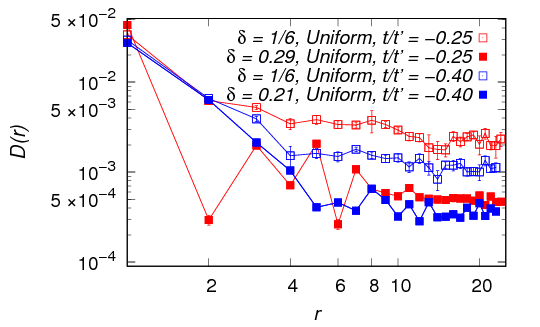}
\centering
\caption{Comparison of the pair-pair correlations $D(r)$ for $t^\prime/t=-0.25$ and $t'/t=-0.4$, as a function of $r$ on a log-log scale. 
Data are reported for the uniform but not optimal superconducting states at $\delta=1/6$ (empty red squares at $t'/t=-0.25$ and empty blue squares at $t'/t=-0.4$) and for the two uniform states at the critical dopings $\delta_c=0.29$ for $t'/t=-0.25$ (full red squares) and $\delta_c=0.21$ for $t'/t=-0.4$ (full blue squares).}
\label{fig: corrSC_mag}
\end{figure}

To conclude the discussion, we compare the magnitude of the superconducting correlations for the two different values of $t'/t$. 
The curves are plotted in Fig.~\ref{fig: corrSC_mag}. We observe how, when $|t^\prime/t|$ is larger, 
superconductivity for the (nonoptimal) uniform state at $\delta=1/6$ is slightly suppressed. This is equally true for the uniform states at the critical dopings $\delta_c$ 
where the uniform state is restored: The effect of a larger value of $|t^\prime/t|$ in suppressing superconductivity is clearly visible, 
since superconducting correlations at the critical dopings are smaller at $t'/t=-0.4$ than at $t'/t=-0.25$, even if $\delta_c$ is definitely 
smaller in the first case.

%%%%%%%%%%%%%%%%%%%%%%%%%%%%%%%%%%%%%%%%%%%%%%%
%%%%%%%%%%%%%%%%%%%%%%%%%%%%%%%%%%%%

\section{Conclusions}

\begin{figure*}
\includegraphics[width=0.8\textwidth]{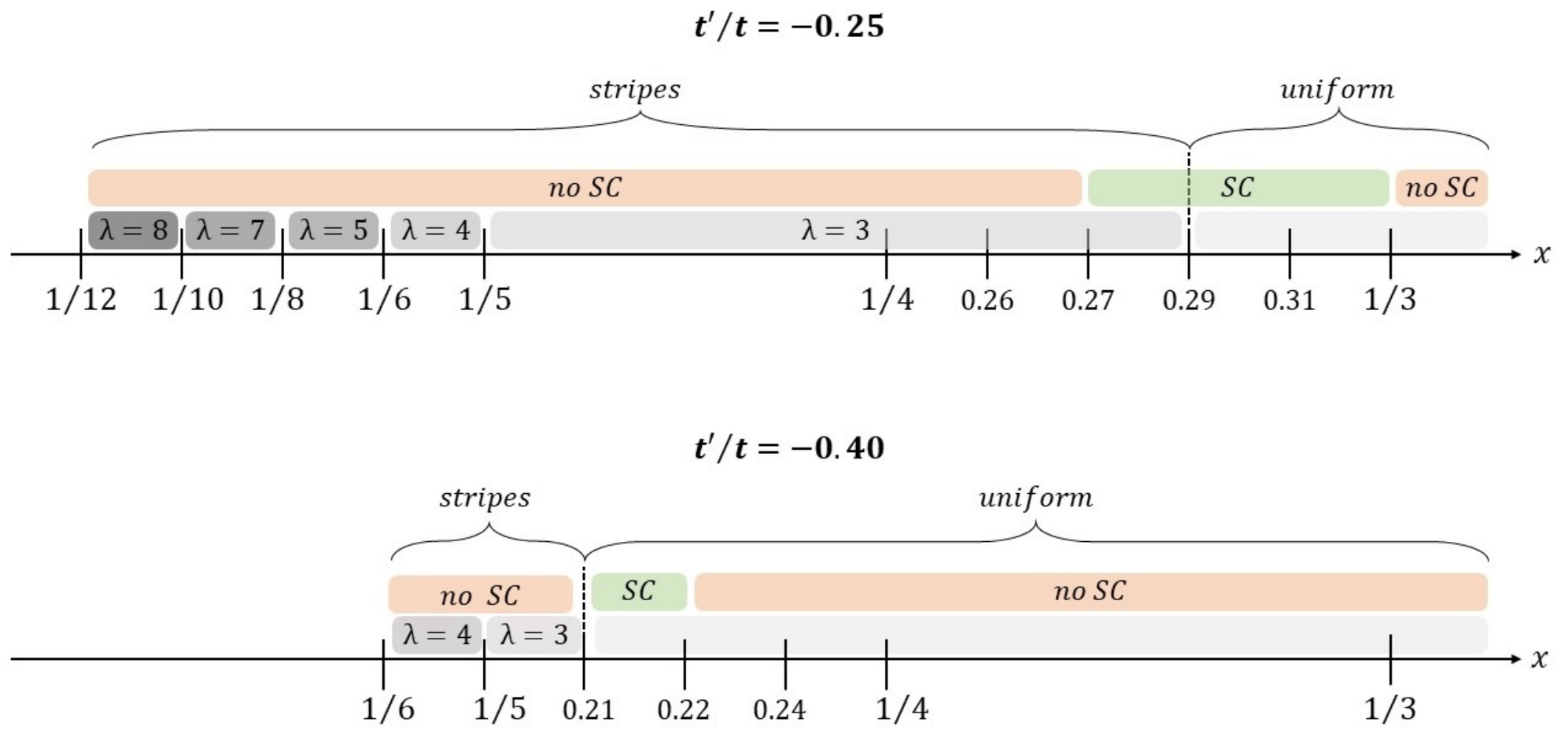}
\centering
\caption{Phase diagram collecting the optimal states as a function of doping $\delta$, for two values of $t'/t$. 
The symbol SC denotes superconductivity, while $\lambda$ indicates the wavelength of the optimal stripe. 
}
\label{fig: final_PD}
\end{figure*}

We have explored the consequences of increasing hole doping on the
instauration of stripe order, superconductivity and their reciprocal interplay, for two prototypical values of $t'/t$.
All the main results of the present work are collected and summarized by the final phase diagram reported in Fig.~\ref{fig: final_PD}. 
Superconductivity is considered to be present when the average of $D(r)$ over the last 10 distances is above a threshold value of $3\times10^{-4}$
~\footnote{This threshold value is chosen arbitrarily, but the goal is to show evidences for some residual superconductivity in between the 
striped states and the uniform nonsuperconducting metal at large doping.}.

%%%%%%%%%%%%%%%%%%%%%%%%%%%%%%%%%%%%%%%%

By looking for the optimal state for different values of the hole-doping $\delta$, 
we found that stripes are present over a broad range of doping values, as they are energetically favorable in comparison to the uniform state. 
Remarkably, site and bond-centered stripes have been found to be essentially degenerate in energy, suggesting that there is no relevant difference between the two configurations. 
Upon increasing $\delta$, the wavelength of the stripes shrinks until eventually the uniform state is restored. A larger $|t^\prime/t|$ is associated to a 
shrinking of the wavelength $\lambda$ and leads to a 
faster dissolution of the stripes, with the uniform state being the optimal one at a smaller value of $\delta$. 

The coexistence of superconductivity and stripe order is addressed by looking at the pair-pair superconducting correlations $D(r)$. 
For both values of $t^\prime/t$, superconductivity is found to be suppressed whenever stripes (no matter their metallic or insulating nature) are present, 
suggesting that the two phenomena interfere with each other. There is then a small interval in $\delta$ among which the hole-doping is strong enough to restore 
the uniform state but not too strong to completely suppress superconductivity. Furthermore, our results show that, at $t'/t=-0.4$ all superconducting correlations 
are weaker than at $t'/t=-0.25$, even if stripes melt at a smaller value of $\delta$.

In conclusion, our results confirm that the phase diagram of the Hubbard model is dominated by stripe states, possibly 
overestimating this phase with respect to superconductivity, when connected to the cuprate physics.

\ack

Computational resources were provided by HPC@POLITO (http://www.
hpc.polito.it). 

\section*{References}

\bibliography{bibliography}

\end{document}